 \documentclass[preprint2]{aastex}

\usepackage{graphicx}
\usepackage{txfonts}

\renewcommand{\ion}[2]{#1\,{\sc #2}}

\shorttitle{The magnetic field of $\zeta$\,Puppis}
\shortauthors{Hubrig et al.}

\begin{document}


\title{
The first spectropolarimetric monitoring of the peculiar O4\,Ief supergiant $\zeta$\,Puppis
}






\author{S.~Hubrig}
\affil{Leibniz-Institut f\"ur Astrophysik Potsdam (AIP), An der Sternwarte 16, 14482 Potsdam, Germany}
\email{shubrig@aip.de}

\author{A.~Kholtygin}
\affil{Astronomical Institute, St.~Petersburg State University, Universitetski pr.~28, 198504, St.~Petersburg, Russia}

\author{I.~Ilyin}
\affil{Leibniz-Institut f\"ur Astrophysik Potsdam (AIP), An der Sternwarte 16, 14482 Potsdam, Germany}

\author{M.~Sch\"oller}
\affil{European Southern Observatory, Karl-Schwarzschild-Str.~2, 85748~Garching, Germany}

\and

\author{L.~M.~Oskinova}
\affil{Universit\"at Potsdam, Institut f\"ur Physik und Astronomie, 14476~Potsdam, Germany}




\begin{abstract}
The origin of the magnetic field in massive O-type stars is still under debate.
To model the physical processes responsible for the generation of O star magnetic fields,
it is important to understand whether correlations between the presence of a magnetic field and 
stellar evolutionary state, rotation velocity, kinematical status, and surface composition can be identified.
The O4\,Ief supergiant $\zeta$\,Pup is a fast rotator and a runaway star, which may be a product 
of a past binary interaction, possibly having had an encounter with the cluster Trumper~10 some 2\,Myr ago.
The currently available observational 
material suggests that certain observed phenomena in this star may be related to the presence of a magnetic field. 
We acquired spectropolarimetric observations of $\zeta$\,Pup
with FORS\,2 mounted  on the 8-m Antu telescope of the VLT to investigate if a magnetic
field is indeed present in this star. 
We show that many spectral lines are highly variable and probably vary with the recently
detected period of 1.78\,d.
No magnetic field is detected in $\zeta$\,Pup, as no magnetic field measurement has a 
significance level higher than 2.4$\sigma$.
Still, we studied the probability of a single sinusoidal explaining
the variation of the longitudinal magnetic field measurements. 
\end{abstract}




\section{Introduction}

Magnetic fields in massive stars have fundamental effects on the stellar evolution and rotation, 
and on the structure, dynamics and heating of their radiatively-driven winds.
To identify and to model the physical processes responsible for the generation of their magnetic fields,
it is important to establish whether magnetic fields can also be detected in massive stars that are fast rotators
and have runaway status.
Recent detections of strong magnetic fields in very fast rotating early-B 
type stars indicate that the spindown timescale via magnetic braking can be much longer than the estimated 
age of these targets (e.g.\ \citealt{rivinius2013}).
Furthermore, current studies of their kinematical status identified a number of magnetic O and Of?p stars
as candidate runaway stars (e.g.\ \citealt{Hubrig2011c}).
Increasing the known number of magnetic objects with extreme rotation, which are probably 
products of a past binary interaction, is important to understand the magnetic field origin in massive stars.

A few years ago, we presented the first spectropolarimetric observations of $\zeta$\,Oph \citep{Hubrig2011a},
obtained with FORS\,1 (FOcal Reducer low dispersion Spectrograph; \citealt{Appenzeller1998}) 
mounted on the 8-m Kueyen telescope of the Very Large Telescope (VLT).
The star $\zeta$\,Ophiuchi (=HD\,149757) of spectral 
type O9.5V is a well-known rapidly rotating runaway star, rotating almost at break-up
velocity with $v\,\sin\,i=400$\,km\,s$^{-1}$ \citep{Kambe1993}. 
Later, additional nine FORS\,2 spectropolarimetric observations of $\zeta$\,Oph were obtained over four consecutive nights.
The analysis of the FORS\,2 observations showed the presence of a weak magnetic field
with a reversal of polarity \citep{Hubrig2013} and an amplitude of about 100\,G. 
The resulting periodogram for the  magnetic  field  measurements  using all available lines showed
a dominating peak corresponding to a period of about  1.3\,d, which is 
roughly double the period of 0.643\,d determined by \citet{Pollmann12}, who studied the variation of the equivalent
width of the \ion{He}{i}~6678 line.
\citet{David2014} used ESPaDOnS and NARVAL observations 
of $\zeta$\,Oph in 2011/12 with much larger uncertainties
-- typically around 350\,G and between 100 and 800\,G --
and concluded that the upper limit of the dipole field strength is about 224\,G
and the maximum possible rotation period 1.1\,d. High-resolution spectropolarimetric results are thus in 
contradiction with the FORS\,2 observations.

The O4\,Ief supergiant $\zeta$\,Pup (=HD\,66811) is the brightest O-type star in the sky with a magnitude $V=2.25$.
Similar to  $\zeta$\,Ophiuchi, it rotates rather fast with $v\,\sin\,i=208$\,km\,s$^{-1}$ \citep{conti1977},
and is a runaway star, 
which may be a product of a past binary interaction, possibly having had an encounter
with the cluster Trumper\,10 some 2\,Myr ago \citep{Hoogerwerf2001,Schilbach08}.
The most recent periodicity analysis was presented by \citet{How2014}, who 
determined $P = 1.780938\pm 0.000093$\,d using optical photometry obtained with the SMEI (Solar Mass Ejection Imager) 
instrument on the Coriolis satellite between 2003 and 2006.
On the other hand, numerous studies in the past that used 
observations in a different part of the electromagnetic spectrum,
reported various periodic or cyclical signals, from a couple of hours to about 5\,days.
An overview of these studies is presented in the work of \citet{How2014}.
The authors report that no evidence for persistent coherent signals with semi-amplitudes in excess of about 2~mmag on 
any of the time-scales previously reported in the literature was detected in their data. In particular, there was 
no evidence for a signature of a rotation period of about 5\,d. Noteworthy, the period of 1.78\,d was also 
recently unambiguously identified in observations of the BRITE (BRIght-star Target Explorer) satellites 
(Ramiaramanantsoa, priv.\ comm.).

The peculiar behaviour, in particular the appearance of a double-peaked \ion{He}{ii} $\lambda$4686 emission line
in the spectra of $\zeta$\,Pup was already discussed over the last
decades (e.g.\ \citealt{conti1974}).
\citet{har2000}, \citet{har2002},  and  \citet{vink2009}  
reported the presence of complex 
linear polarization effects across the H$\alpha$ line profiles of $\zeta$\,Pup and suggested that they can be
attributed to the impact of the rapid rotation of the star on its winds.
\citet{moffat1981} detected profile variability with a period of 5.075\,d
in the depth of the nearly central absorption reversal of the P\,Cygni type H$\alpha$ profile
in spectrophotometric observations of $\zeta$\,Pup
and proposed that the winds of Oef stars might be shaped by a dipolar stellar magnetic field. 
Searches for the presence of a magnetic field in $\zeta$\,Pup were reported by \citet{ChesneauMoffat2002} and
\citet{David2014}.
\citet{ChesneauMoffat2002} used CASPEC observations on four consecutive nights,
covering time intervals between 30 and 76\,min.
The measurements were based on four He lines placed well within
the CASPEC orders and not sensitive to defects in the normalization procedure.
Although a change of field polarity was indicated on every second night,
the measurement accuracies were low, between 200 and 360\,G.
\citet{David2014} obtained 30 individual measurements over 1.9\,h with ESPaDOnS in a single night.
From these observations, the authors estimated an upper limit on the dipolar field strength of 121\,G. 
On the other hand, high-resolution spectropolarimetric observations are not
especially appropriate for the detection of weak magnetic fields in broad-lined stars.
Some broad spectral lines like the \ion{He}{ii}~4686 line or the close-by blend of \ion{N}{iii} lines extend 
over adjacent orders, so that it is necessary to adopt the order 
shapes to get the best continuum normalization.
No continuum normalization is required in FORS\,2 spectropolarimetric observations.

Furthermore, it is of importance that \citet{baade1986} and  \citet{Reid1996} reported the presence of 
velocity-resolved structure in the photospheric absorption lines of $\zeta$\,Pup, with a possible 8.5-h periodicity. 
Their observations showed characteristic blue-to-red 
migration of bumps and dips in the absorption-line profiles, suggesting non-radial pulsation as the underlying physical mechanism.
\citet{baade1986} also mentioned that every 0.178\,d a new component develops in the blue wings of the \ion{C}{iv}~5801, 5812 lines
and it takes twice the amount of time for the feature to traverse the entire profile.
Notably, \citet{Eversberg1998} report on the presence of variable substructures 
on a time scale of 10-20\,min, likely representing the stochastic manifestation of turbulent clumps 
propagating outward with the wind.
These studies underline the possible impact of short-time variability on the magnetic field 
measurements, which can be quite significant in supergiants (see Fig.~\ref{fig:same} and e.g.\ \citealt{Hubrig2014}).

Since the information on the presence of a magnetic field in $\zeta$\,Pup is rather limited,
we decided to carry out spectropolarimetric observations randomly distributed over a few weeks using 
the VLT instrument FORS\,2 in service mode.
In this paper, we present the results of our FORS\,2 spectropolarimetric monitoring of $\zeta$\,Pup and discuss the observed 
magnetic and spectral variability assuming the oblique rotator interpretation.


\section{Magnetic field measurements}

Thirteen spectropolarimetric observations of $\zeta$\,Pup were carried
out from 2013 October 5 to 2013 December 23 in service mode at the European
Southern Observatory using FORS\,2
\citep{Appenzeller1998} mounted on the 8-m Antu telescope of the VLT.
This multi-mode instrument is equipped with polarisation analysing optics
comprising super-achromatic half-wave and quarter-wave phase retarder plates,
and a Wollaston prism with a beam divergence of 22$\arcsec$ in standard
resolution mode. 
We used the GRISM 600B and the narrowest available slit width
of 0$\farcs$4 to obtain a spectral resolving power of $R\sim2000$.
The observed spectral range from 3250 to 6215\,\AA{} includes all Balmer lines,
apart from H$\alpha$, down to H14, and numerous helium lines, up to 15.
For the observations, we used a non-standard readout mode with low 
gain (200kHz,1$\times$1,low), which provides a broader dynamic range, hence 
allowed us to reach a higher signal-to-noise ratio (SNR) in the individual spectra.

A first description of the assessment of the longitudinal magnetic field
measurements using FORS\,1/2 spectropolarimetric observations was presented 
in our previous work (e.g.\ \citealt{Hubrig2004a,Hubrig2004b}, 
and references therein).
To minimize the cross-talk effect,
and to cancel errors from 
different transmission properties of the two polarised beams,
a sequence of subexposures at the retarder
position angles
$-$45$^{\circ}$$+$45$^{\circ}$,
$+$45$^{\circ}$$-$45$^{\circ}$,
$-$45$^{\circ}$$+$45$^{\circ}$,
etc.\ is usually executed during observations. Moreover, the reversal of the quarter wave 
plate compensates for fixed errors in the relative wavelength calibrations of the two
polarised spectra.
According to the FORS User Manual, the $V/I$ spectrum is calculated using:
\begin{equation}
\frac{V}{I} = \frac{1}{2} \left\{ 
\left( \frac{f^{\rm o} - f^{\rm e}}{f^{\rm o} + f^{\rm e}} \right)_{-45^{\circ}} -
\left( \frac{f^{\rm o} - f^{\rm e}}{f^{\rm o} + f^{\rm e}} \right)_{+45^{\circ}} \right\}
\end{equation}
where $+45^{\circ}$ and $-45^{\circ}$ indicate the position angle of the
retarder waveplate and $f^{\rm o}$ and $f^{\rm e}$ are the ordinary and
extraordinary beams, respectively.  Rectification of the $V/I$ spectra was
performed in the way described by \citet{Hubrig2014}.
Null profiles, $N$, are calculated as pairwise differences from all available 
$V$ profiles.  From these, 3$\sigma$-outliers are identified and used to clip 
the $V$ profiles.  This removes spurious signals, which mostly come from cosmic
rays, and also reduces the noise. A full description of the updated data 
reduction and analysis will be presented in a separate paper (Sch\"oller et 
al., in preparation). Due to the brightness of the star, the exposure time for each subexposure 
accounted for 0.25\,s.

The mean longitudinal magnetic field, $\left< B_{\rm z}\right>$, is 
measured on the rectified and clipped spectra based on the relation
\begin{eqnarray} 
\frac{V}{I} = -\frac{g_{\rm eff}\, e \,\lambda^2}{4\pi\,m_{\rm e}\,c^2}\,
\frac{1}{I}\,\frac{{\rm d}I}{{\rm d}\lambda} \left<B_{\rm z}\right>\, ,
\label{eqn:vi}
\end{eqnarray} 

\noindent 
where $V$ is the Stokes parameter that measures the circular polarization, $I$
is the intensity in the unpolarized spectrum, $g_{\rm eff}$ is the effective
Land\'e factor, $e$ is the electron charge, $\lambda$ is the wavelength,
$m_{\rm e}$ is the electron mass, $c$ is the speed of light, 
${{\rm d}I/{\rm d}\lambda}$ is the wavelength derivative of Stokes~$I$, and 
$\left<B_{\rm z}\right>$ is the mean longitudinal (line-of-sight) magnetic field (e.g.\ \citealt{angel1970}).

The longitudinal magnetic field was measured in three ways: using the entire spectrum
including all available lines, excluding lines in emission, and using exclusively hydrogen lines.
The feasibility of longitudinal magnetic field measurements in massive stars 
using low-resolution spectropolarimetric observations was demonstrated by previous studies of O and B-type stars
(e.g., \citealt{Hubrig2006,Hubrig2008,Hubrig2009,Hubrig2011b,Hubrig2013,Hubrig2015a}).
A number of discrepancies in the published measurement accuracies using FORS\,1 data
were reported by \citet{Bagnulo2012},
who used the ESO FORS\,1 pipeline to reduce the FORS\,1 archive.
Some of the reasons for higher error bars
in the study of \citet{Bagnulo2012} are discussed by \citet{Hubrig2016}.
Furthermore, we have carried out Monte Carlo bootstrapping tests. 
These are most often applied with the purpose of deriving robust estimates of standard errors. 
The measurement uncertainties obtained before and after Monte Carlo bootstrapping tests were found to be 
in close agreement, indicating the absence of reduction flaws. 
Since \citet{baade1986} and  \citet{Reid1996} suggested the presence of non-radial 
pulsations in $\zeta$\,Pup, to check the stability 
of the spectral lines along the full sequence of sub-exposures, we have compared 
the profiles of several spectral lines recorded in the parallel beam with the retarder waveplate at $+45^{\circ}$. 
The same was done for spectral lines recorded in the perpendicular beam. 
The line profiles looked identical within the noise and no instability like that found in the A0 supergiant HD\,92207
\citep{Hubrig2014} was found in our FORS\,2 observations.

\begin{table*}
\caption{
Logbook of the FORS\,2 polarimetric observations of $\zeta$\,Pup, including 
the modified Julian date of mid-exposure followed by the
achieved signal-to-noise ratio in the Stokes~$I$ spectrum around 5000\,\AA{},
and the measurements of the mean longitudinal magnetic field using the 
Monte Carlo bootstrapping test, for all lines, excluding the lines in emission, and for the hydrogen lines.
In the last columns, we present the results of our measurements using the null spectra and the phases calculated
relative to the zero phase corresponding to the negative field extremum at MJD56570.2611.
All quoted errors are 1$\sigma$ uncertainties.
}
\label{tab:log_meas}
\centering
\begin{tabular}{lrr@{$\pm$}rr@{$\pm$}rr@{$\pm$}rr@{$\pm$}rr}
\hline
\hline
\multicolumn{1}{c}{MJD} &
\multicolumn{1}{c}{SNR$_{5000}$} &
\multicolumn{2}{c}{$\left< B_{\rm z}\right>_{\rm all}$} &
\multicolumn{2}{c}{$\left< B_{\rm z}\right>_{\rm no\,emiss}$} &
\multicolumn{2}{c}{$\left< B_{\rm z}\right>_{\rm hydr}$} &
\multicolumn{2}{c}{$\left< B_{\rm z}\right>_{\rm N}$} &
\multicolumn{1}{c}{Phase}\\
 &
 &
 \multicolumn{2}{c}{[G]} &
 \multicolumn{2}{c}{[G]}  &
 \multicolumn{2}{c}{[G]}  &
 \multicolumn{2}{c}{[G]} &
 \\
\hline
      56570.3265&  934 &$-$420  & 178 & $-$441 & 181 & $-$372 &259&$-$32 &190& 0.037 \\ 
      56622.3478& 2396 &    25  &  61 &    61  & 65  &    99  &139&    3 & 71& 0.247\\  
      56629.2268& 2628 &     4  &  67 &  $-$12 & 71  & $-$24  &110& $-$2 & 60& 0.110  \\  
      56630.2214& 4013 &    110 &  60 &    171 & 77  &   147  &106&   33 & 48& 0.668 \\  
      56631.2471& 3428 &      2 &  44 &     53 & 46  &    26  & 91&  $-$6& 55& 0.244\\  
      56632.1923& 2781 &  $-$12 &  77 &  $-$44 & 67  & $-$77  &104&  $-$27& 88&0.775\\  
      56643.1478& 2137 &  $-$67 &  60 &  $-$78 & 72  & $-$50  &149&     74& 86& 0.926\\  
      56644.2183& 2337 &     22 &  91 &     18 & 101 &    62  &136&  $-$33& 93& 0.527\\  
      56646.0849& 2373 &    132 &  72 &    148 &  78 &    122 &138&  $-$59& 78& 0.575\\  
      56647.1013& 2222 &     34 &  91 &      4 &  94 & $-$110 &134&  $-$31& 106& 0.146 \\  
      56647.2830& 2365 &  $-$17 &  47 &      1 &  57 &     68 &146&   $-$9& 60& 0.248 \\  
      56648.2853& 2371 &     34 &  82 &     47 & 101 &     51 &134&  $-$17& 102& 0.811\\  
      56649.1691& 2175 &     98 &  97 &     92 & 110 &    142 &137&     17& 98& 0.307 \\  
\hline
\end{tabular}
\end{table*}

The results of our magnetic field measurements, those for absorption and emission lines combined,
absorption lines alone, and hydrogen lines are presented in 
Table~\ref{tab:log_meas}, where we also collect information about the modified Julian date 
for the middle of the exposure, 
the achieved peak signal-to-noise ratio in the Stokes~$I$ spectrum close to 5000\,\AA{},
the measurements obtained using null spectra for the set of all lines, 
and the rotation phase assuming a rotation period of 1.780938\,d. The results using null 
spectra for the other sets of lines are similar.
The first measurement in Table~\ref{tab:log_meas} shows the strongest longitudinal
magnetic field of about $-400$\,G.
However, the accuracy is very low due to the low SNR obtained for this observation.
In addition, the time interval between this first measurement and the second is
the largest, amounting to almost 52\,d, while all other measurements were obtained
within 27\,d.

\section{Magnetic and spectral variability}

\begin{figure}
\centering
\includegraphics[width=0.40\textwidth]{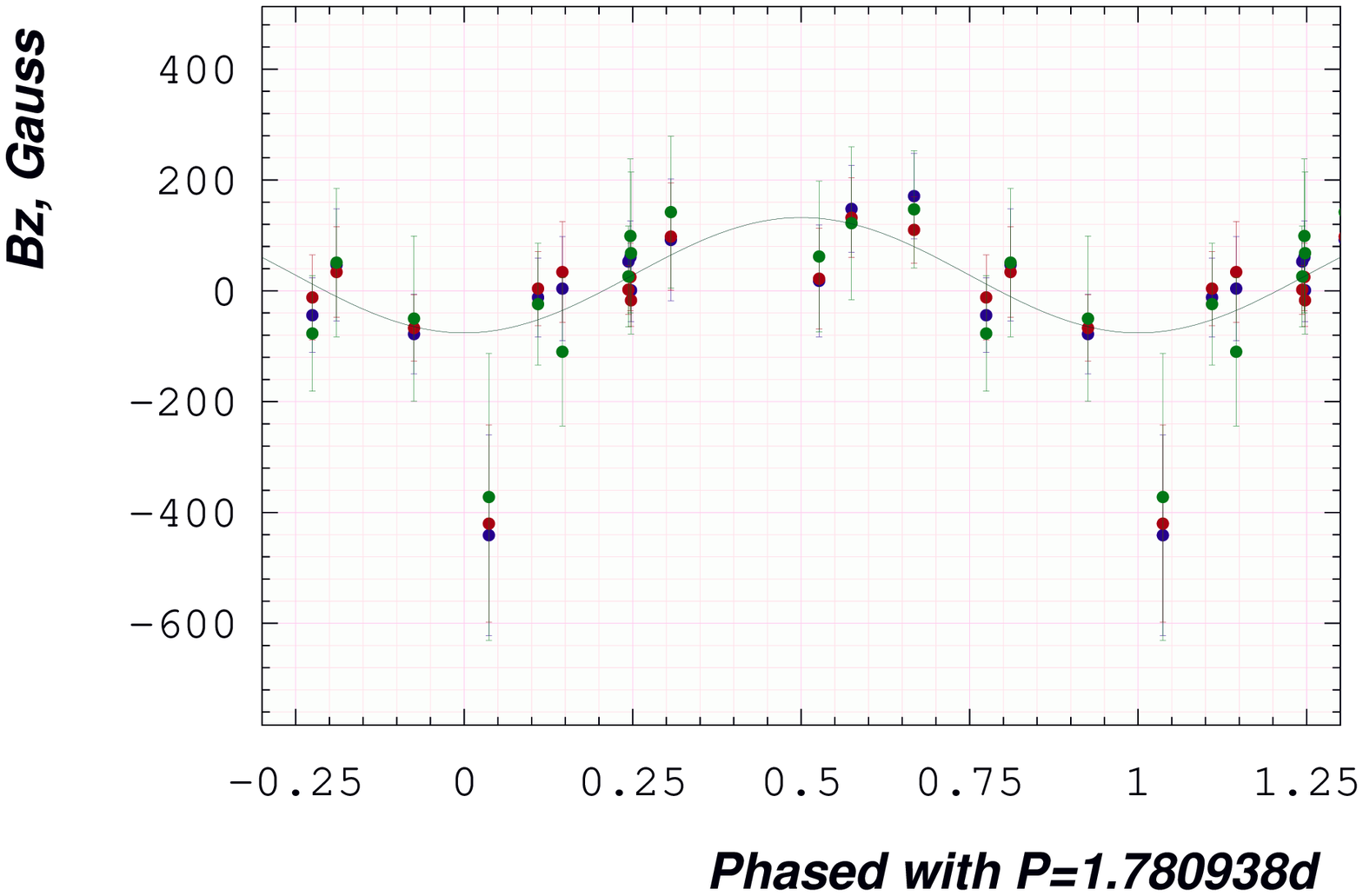}
\caption{
Longitudinal magnetic field variation of $\zeta$\,Pup phased with the 1.78\,d 
period reported by \citet{How2014}.
Measurements of the magnetic field using absorption and emission lines are presented by red circles,
those without emission lines by blue circles and those for hydrogen lines by green circles.
The solid line represents a fit to the data with a mean value for the magnetic field of
$\overline{\left<B_{\rm z}\right>} = 28.2\pm10.5$\,G, an amplitude of
$A_{\left<B_{\rm z}\right>} = 104.0\pm19.2$\,G, and a root mean square value of 62.2\,G.
For the presented fit, we assumed a zero phase corresponding to the negative field extremum at MJD56570.2611.
}
\label{fig:rot}
\end{figure}

\begin{figure}
\centering
\includegraphics[width=0.40\textwidth]{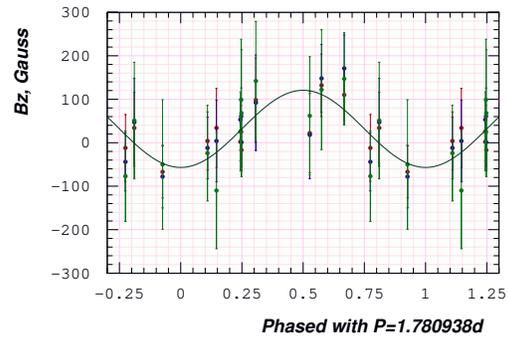}
\caption{
The same as in Fig.~\ref{fig:rot} but excluding the measurement  with the lowest accuracy.
The solid line represents a fit to the data with a mean value for the magnetic field of
$\overline{\left<B_{\rm z}\right>} = 32.3\pm8.2$\,G, an amplitude of
$A_{\left<B_{\rm z}\right>} = 88.4\pm15.2$\,G, and a root mean square value of 46.7\,G.
}
\label{fig:rot1}
\end{figure}

The longitudinal magnetic field variation of 
$\zeta$\,Pup over a period of 1.78\,d is presented in Fig.~\ref{fig:rot},
for absorption and emission lines combined, absorption lines alone, and for hydrogen lines.
The solid line represents a fit to the joined three sets of measurements,
with a mean value of the magnetic field of
$\overline{\left<B_{\rm z}\right>} = 28.2\pm10.5$\,G and an amplitude of
$A_{\left<B_{\rm z}\right>} = 104.0\pm19.2$\,G. 
For the presented fit, we assume a zero phase corresponding to the negative field extremum at MJD56570.2611.
The measurements phased with the period 1.78\,d show a single-wave variation. Noteworthy,  our phasing of the magnetic field
measurements with other periods mentioned in the literature do not display such a sinusoidal variability.

No longitudinal magnetic field measurement shows a 3$\sigma$ detection. The highest significance levels in the 
measurements were achieved at phase 0.037 (at 2.4$\sigma$) for the measurements using the entire spectral range 
and at phase 0.668 (at 2.2$\sigma$) for the measurement carried out excluding emission lines.
Among the 
presented measurements, the measurement obtained at phase 0.037 shows the highest field strength, but 
also the lowest accuracy. As the assigned weights are inversely proportional to 
the squares of the measurement errors in the fitting procedure,
the contribution of this measurement is expected to be by a factor of $\sim$7 lower than that for the other measurements.
To check the likelihood of the obtained sinusoidal fit, we repeated the fitting procedure excluding
this measurement. The result is presented in  Fig.~\ref{fig:rot1},
where the solid line shows the fit to the joined three sets of measurements. 
We obtain a rather similar mean value of the magnetic field 
$\overline{\left<B_{\rm z}\right>} = 32.3\pm8.2$\,G, and a slightly smaller field amplitude 
$A_{\left<B_{\rm z}\right>} = 88.4\pm15.2$\,G. For this fitting model superimposed on the data 
presented in Fig.~\ref{fig:rot1}, we calculate a reduced $\chi^2$ of 0.36. 
The reduced $\chi^2$ value calculated assuming a model in which the longitudinal magnetic
field is constant and equal to 0, is 0.84. 
Thus, the zero field assumption also represents the data well.
Low $\chi^2$ values could point to an overestimation of the errors, which we rule out,
or to fitting noise.

\begin{figure}
\centering
\includegraphics[width=0.45\textwidth]{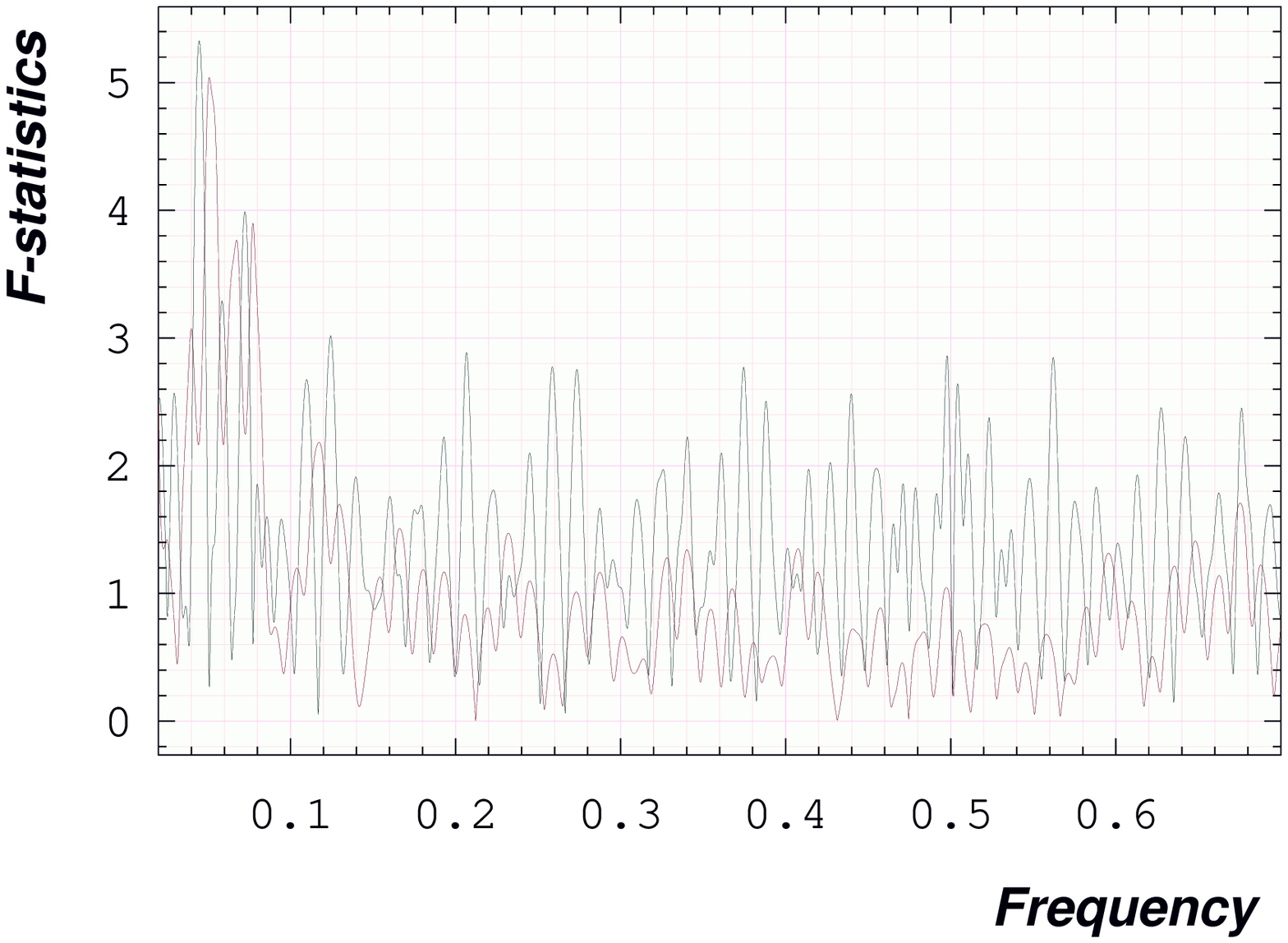}
\caption{
Frequency periodogram (in d$^{-1}$) for the longitudinal magnetic field measurements using the entire spectrum.
The window function is indicated by the red color. 
}
\label{fig:period}
\end{figure}

We note that no periodicity of the magnetic variability is indicated in our frequency periodograms obtained for the 
magnetic field measurements. The result of the frequency analysis performed using a non-linear least squares
fit to the multiple harmonics utilizing the Levenberg-Marquardt
method \citep{press1992} with an optional possibility of prewhitening the trial harmonics is presented in 
Fig.~\ref{fig:period}. To detect 
the most probable period, we calculated the frequency spectrum and for each trial
frequency we performed a statistical F-test of the null hypothesis for the absence of periodicity 
\citep{seber1977}. The resulting F-statistics can be thought of as the total sum including covariances of the ratio 
of harmonic amplitudes to their standard deviations, i.e.\ a signal-to-noise ratio.

\begin{table}
\centering
\caption{
Assumed magnetic field model for $\zeta$\,Pup, including the data point
at phase 0.037 ({\sl left}), and excluding it ({\sl right}).
}
\label{tab:dipvals}
\begin{tabular}{cc|r@{$\pm$}l|r@{$\pm$}l}
\hline
\hline
$\overline{\left< B_{\rm z}\right>}$ & [G]           & 28.2     & 10.5 & 32.3 & 8.2 \\ 
$A_{\left< B_{\rm z}\right>}$        & [G]           & 104.0    & 19.2 & 88.4 & 15.2 \\
$P$                                  & [d]           & \multicolumn{4}{c}{$1.780938 \pm 0.000093$} \\
$R$                                  & [$R_{\odot}$] & \multicolumn{4}{c}{$14.0 \pm 0.4$} \\
$v_{\rm eq}$                         & [km s$^{-1}$] & \multicolumn{4}{c}{$398 \pm 11$} \\
$v$\,sin\,$i$                        & [km s$^{-1}$] & \multicolumn{4}{c}{$208 \pm 5$} \\
$i$                                  & [$^{\circ}$]  & \multicolumn{4}{c}{$31.5 \pm 1.3$} \\
$\beta$                              & [$^{\circ}$]  & 80.6     &  3.9 & 77.4 & 3.8 \\
$B_{\rm d}$                          & [G]           & 681      & 125  & 585  & 93\\
\hline
\end{tabular}
\end{table}

If we assume that the magnetic field variability displayed in  Figs.~\ref{fig:rot} and ~\ref{fig:rot1} is real, then 
the observed  single-wave variation in the longitudinal magnetic field during the stellar rotation cycle would indicate
a dominant dipolar contribution to the magnetic field topology. If the star is an oblique dipole rotator, 
the magnetic dipole axis tilt $\beta$ is constrained by

\begin{equation}
\frac{\left<B_{\rm z}\right>_{\rm max}}{\left<B_{\rm z}\right>_{\rm min}} =\frac{\cos(i + \beta)}{\cos(i - \beta)},
\end{equation}

\noindent
where the inclination
angle $i$ can be derived from considerations of the stellar fundamental parameters \citep{preston1967}.
Using for the stellar radius $R=14\pm0.4\,R_{\odot}$ \citep{Schilbach08}
combined with the period of 1.78\,d, we obtain $v_{\rm eq}=398\pm11$\,km\,s$^{-1}$.
Using $v\,\sin\,i= 208\pm5$\,km\,s$^{-1}$ \citep{conti1977} we obtain the inclination angle $i=31.5\pm1.3^{\circ}$.
In Table~\ref{tab:dipvals}, we show for $\zeta$\,Pup the corresponding parameters
of the possible magnetic field dipole models.
The estimated dipole strengths of $681\pm125$\,G and $585\pm93$\,G are significantly 
higher than the dipole strength value suggested by \citet{David2014},
who determined an upper limit of 121\,G from one single longitudinal magnetic field measurement.
Obviously, greater sensitivity in future
magnetic field observations is necessary to further test the presence of a magnetic field in $\zeta$\,Pup.

\begin{figure}
\centering
\includegraphics[width=0.22\textwidth]{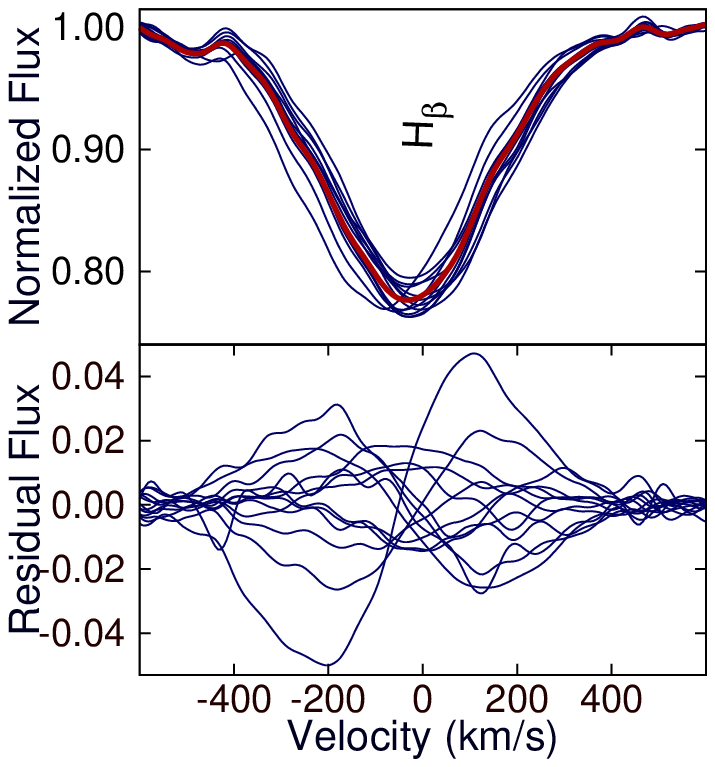}
\includegraphics[width=0.22\textwidth]{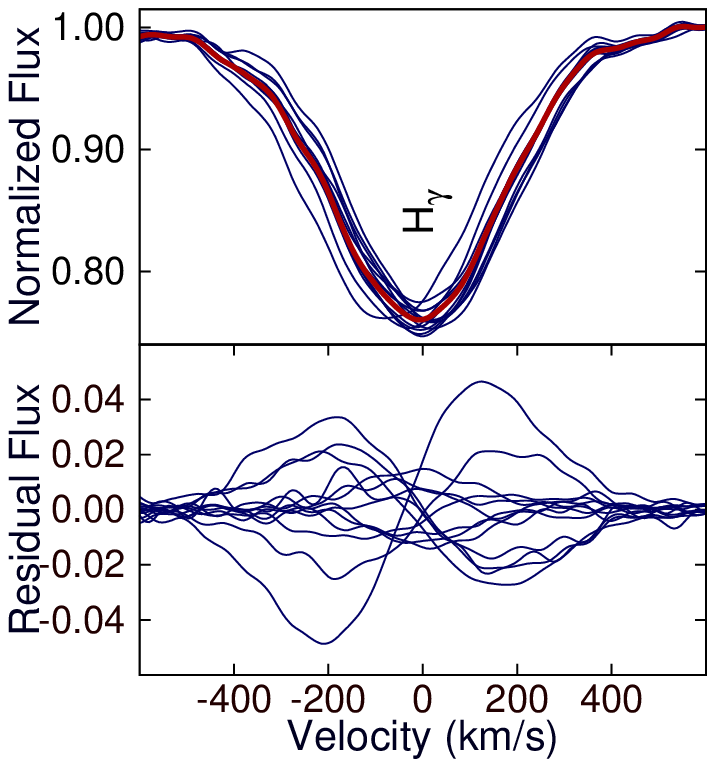}
\includegraphics[width=0.22\textwidth]{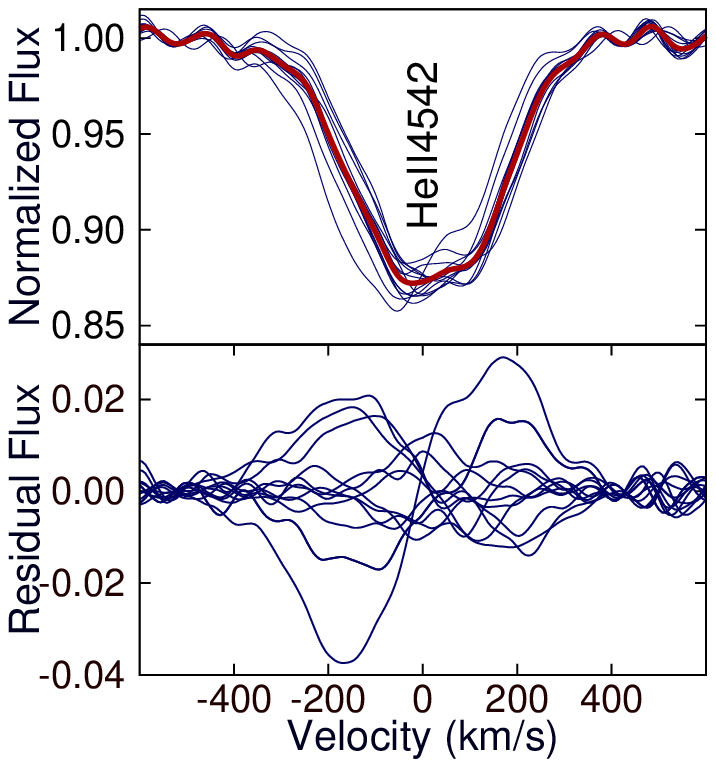}
\includegraphics[width=0.22\textwidth]{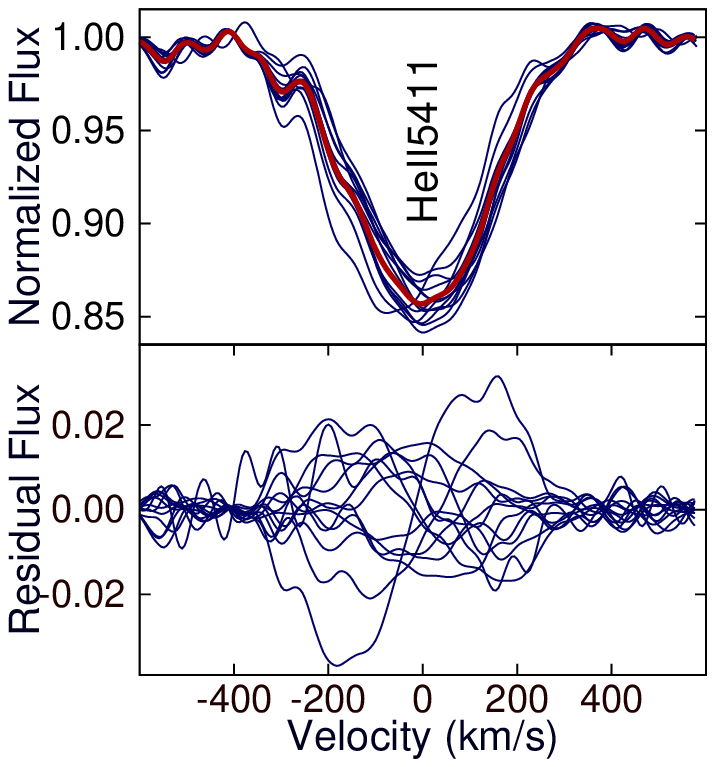}
\includegraphics[width=0.22\textwidth]{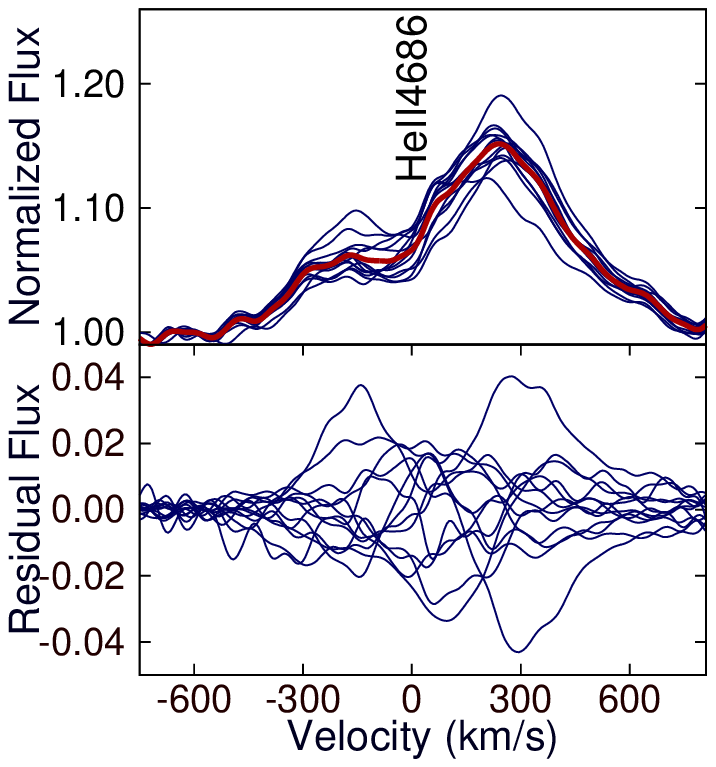}
\caption{
Variability of hydrogen and \ion{He}{ii} lines in FORS\,2 spectra obtained at thirteen different epochs.
The upper panels show overplotted profiles while the lower panels present the differences between individual line profiles 
and the average profiles. The average profile for each line is highlighted in red.
}
\label{fig:overpl}
\end{figure}

\begin{figure}
\centering
\includegraphics[width=0.22\textwidth]{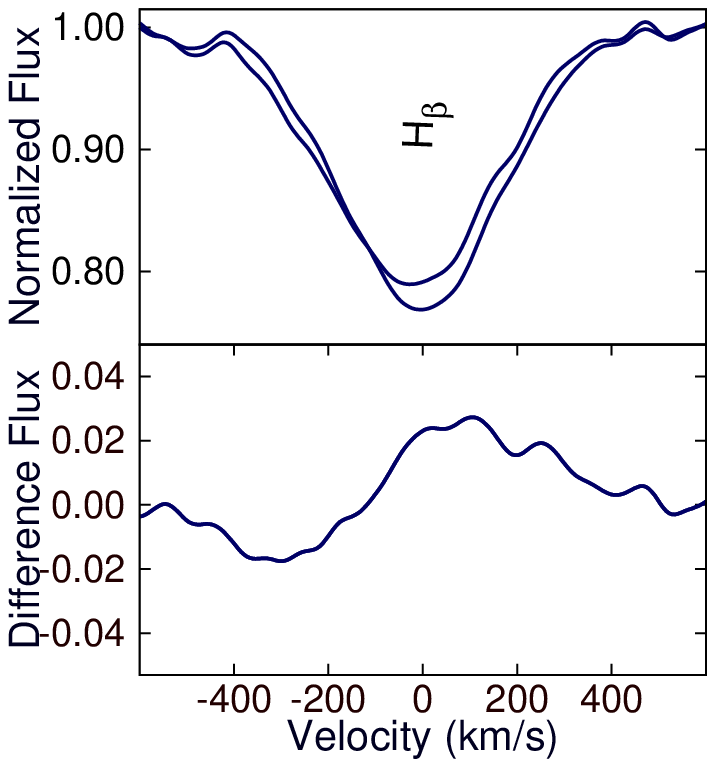}
\includegraphics[width=0.22\textwidth]{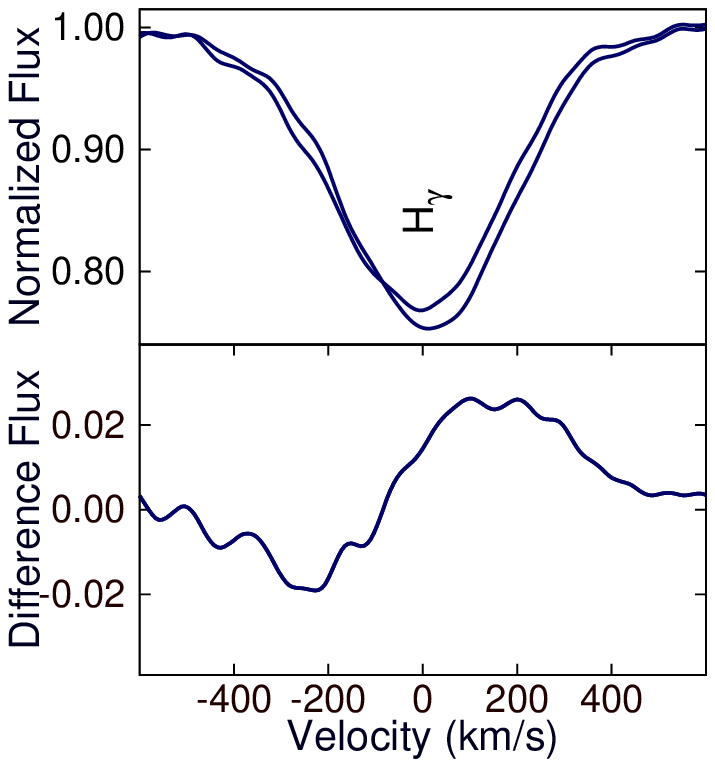}
\includegraphics[width=0.22\textwidth]{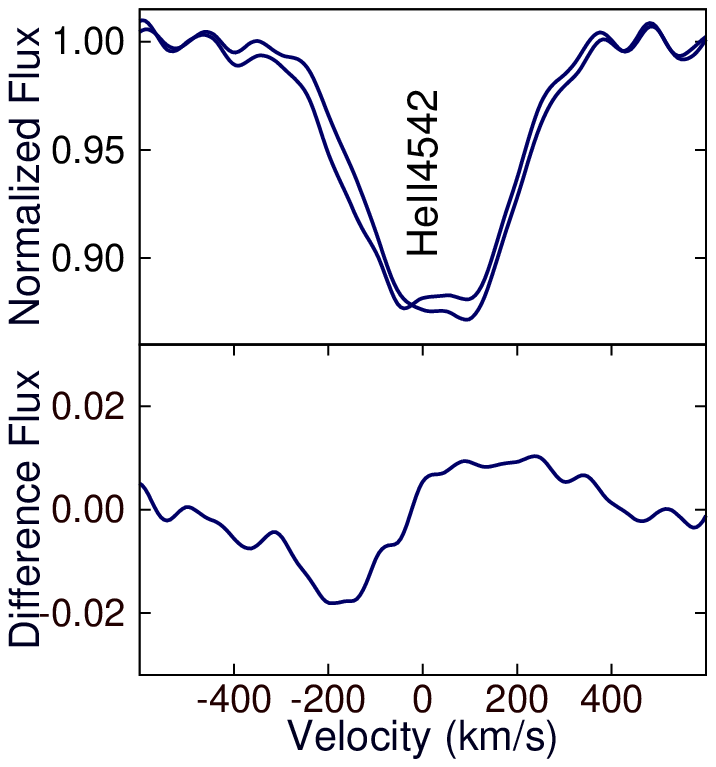}
\includegraphics[width=0.22\textwidth]{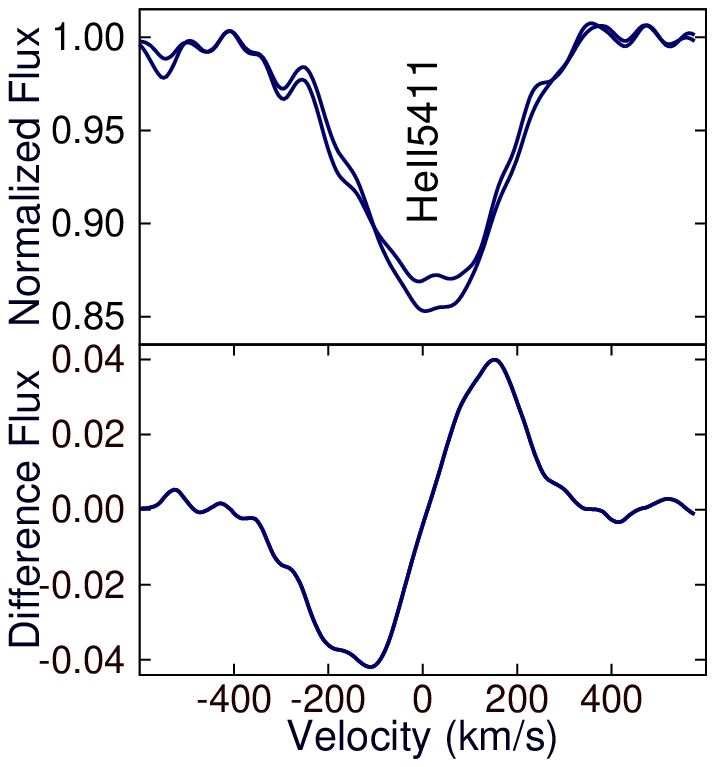}
\includegraphics[width=0.22\textwidth]{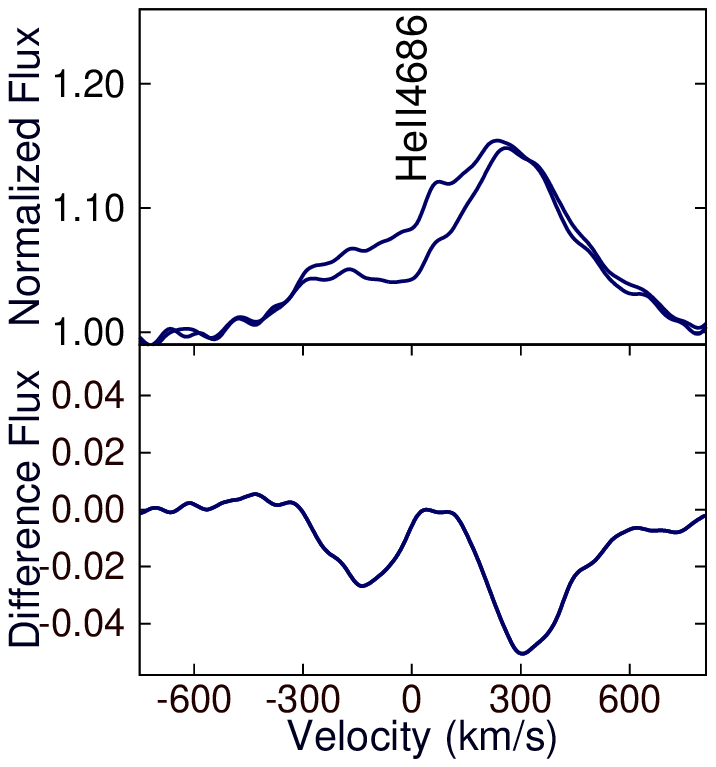}
\caption{
Variability of hydrogen and \ion{He}{ii} lines in FORS\,2 spectra obtained on the same night within  
a time interval of 4.4\,h.
The upper panels show overplotted profiles while the lower panels present the differences between 
the individual line profiles. 
}
\label{fig:same}
\end{figure}

As was already reported in the literature, line profiles belonging to different elements
in $\zeta$\,Pup show strong variability.
As an example, we present in Fig.~\ref{fig:overpl} the overplotted profiles of hydrogen lines H$\beta$ and H$\gamma$
and of the \ion{He}{ii}
lines 4542, 5411, and 4686 and the differences between individual line profiles and the average profiles. 
The largest variability amplitude in the velocity frame and in the line profile depths is detected in hydrogen lines and 
in the \ion{He}{ii}~4686 line. The \ion{He}{ii}~4686 line shows the presence of two components, where the blue component 
becomes stronger in the vicinity of the positive magnetic extremum and slightly stronger again in the vicinity of the negative field
extremum. Similar to the results of \citet{baade1986}, who used the \ion{C}{iv}~5801, 5812 lines, 
the \ion{He}{ii}~4542 line exhibits a clear splitting at certain epochs, while the line \ion{He}{ii}~5411 displays a moderate splitting. 
Although we detect that the intensity of the split components in the \ion{He}{ii}~4542 line 
changes with time, more observations are necessary to assess the character of this variability.
\citet{baade1986}
interpreted the detected variations as non-radial oscillations with a period of about 0.356\,d.
A rather strong spectral variability is also detected on a short time scale: two FORS\,2 observations of $\zeta$\,Pup 
were obtained on the same night on 2013 December 21 within a time interval of only of 4.4\,h.
In Fig.\ref{fig:same},
we present for the same lines as in Fig.~\ref{fig:overpl} the differences between the observations
obtained within the same night. 

\begin{figure}
\centering
\includegraphics[width=0.22\textwidth]{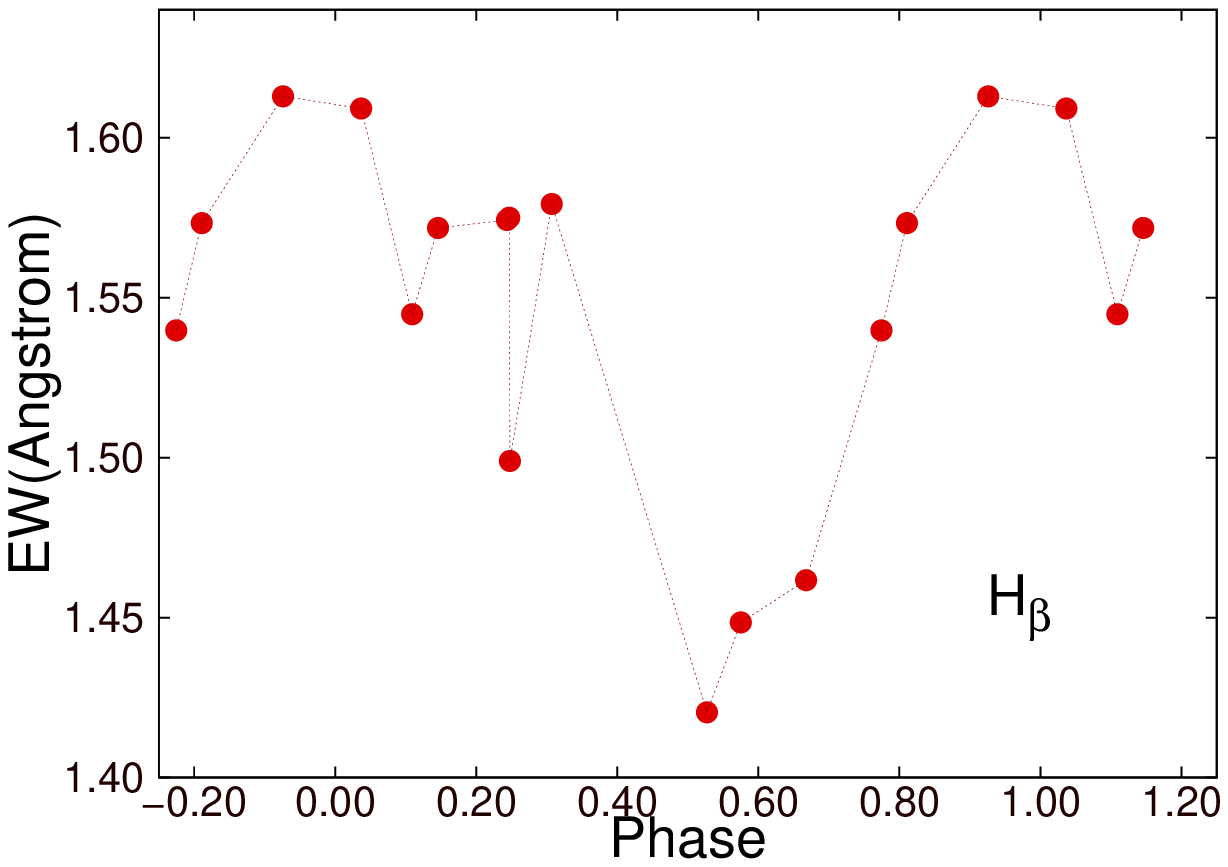}
\includegraphics[width=0.22\textwidth]{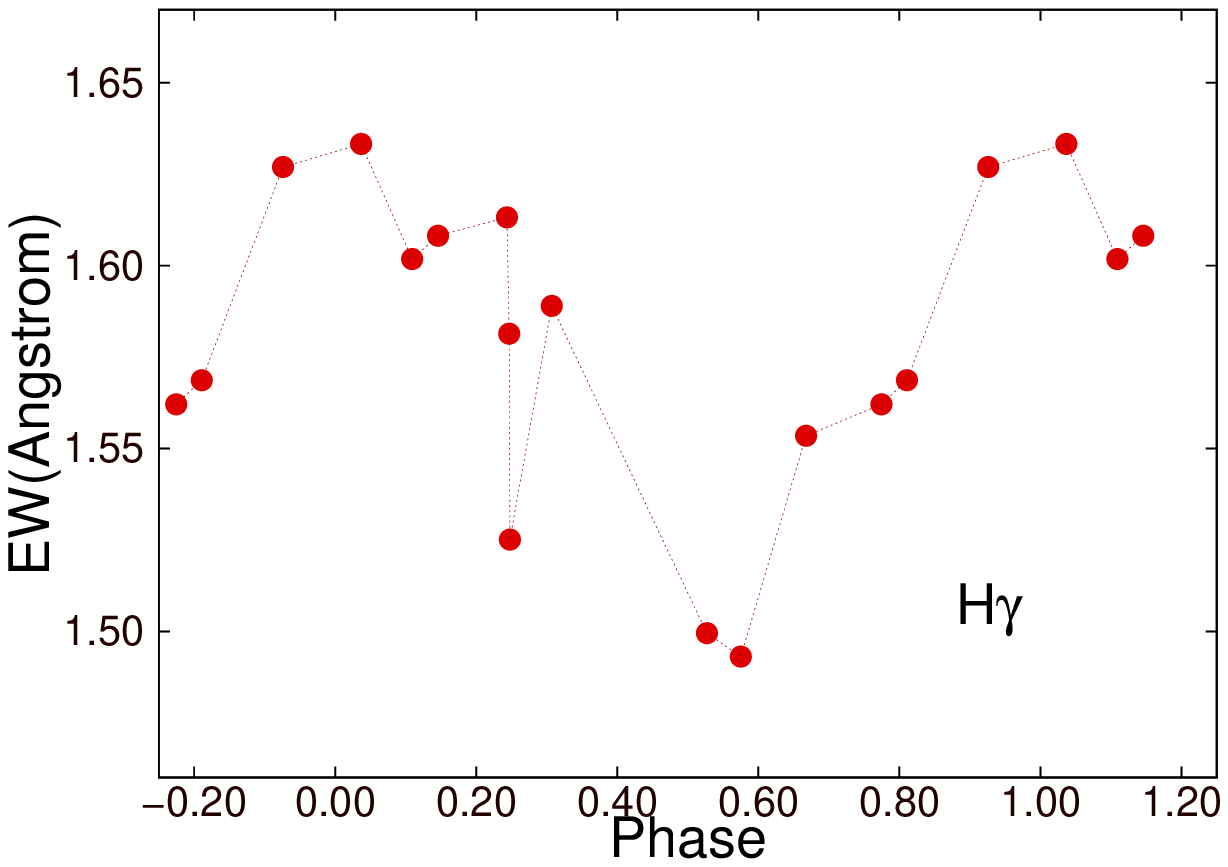}
\includegraphics[width=0.22\textwidth]{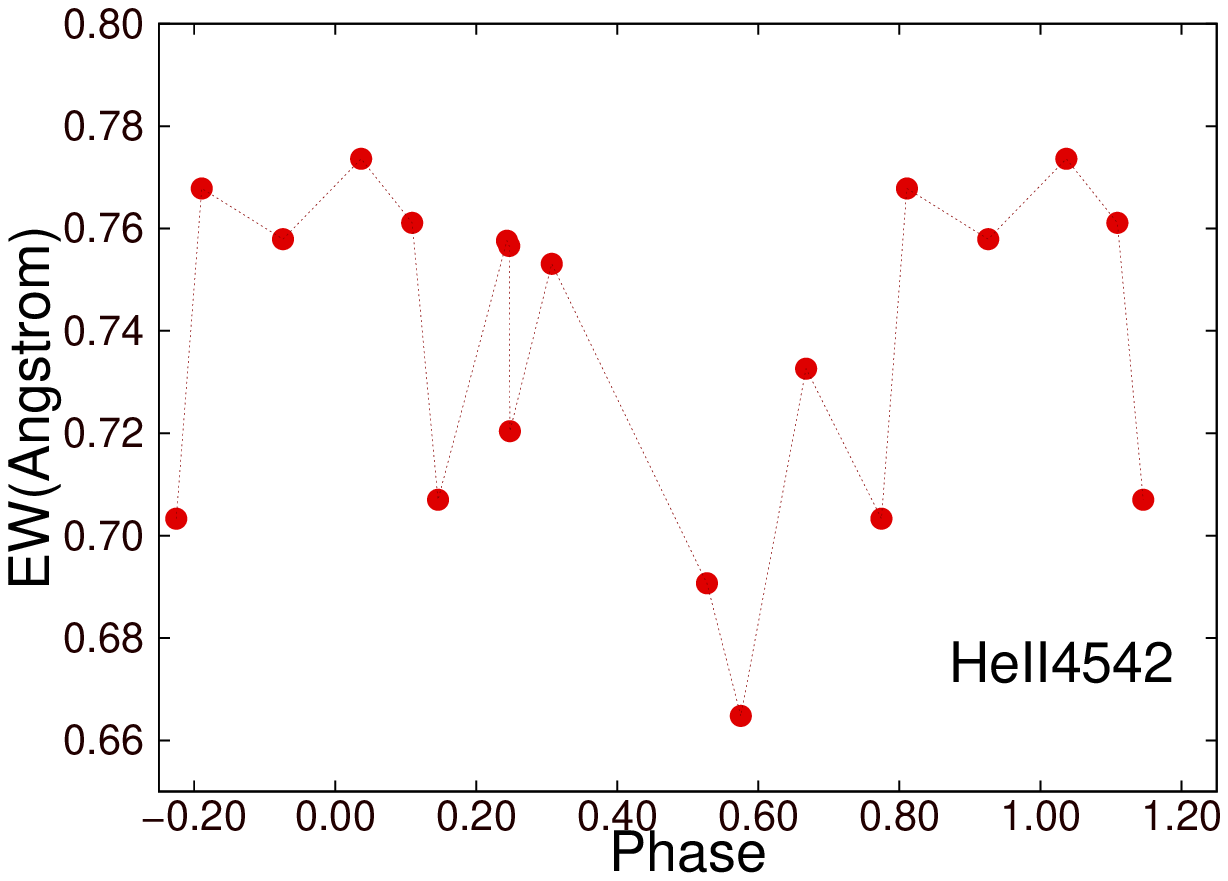}
\includegraphics[width=0.22\textwidth]{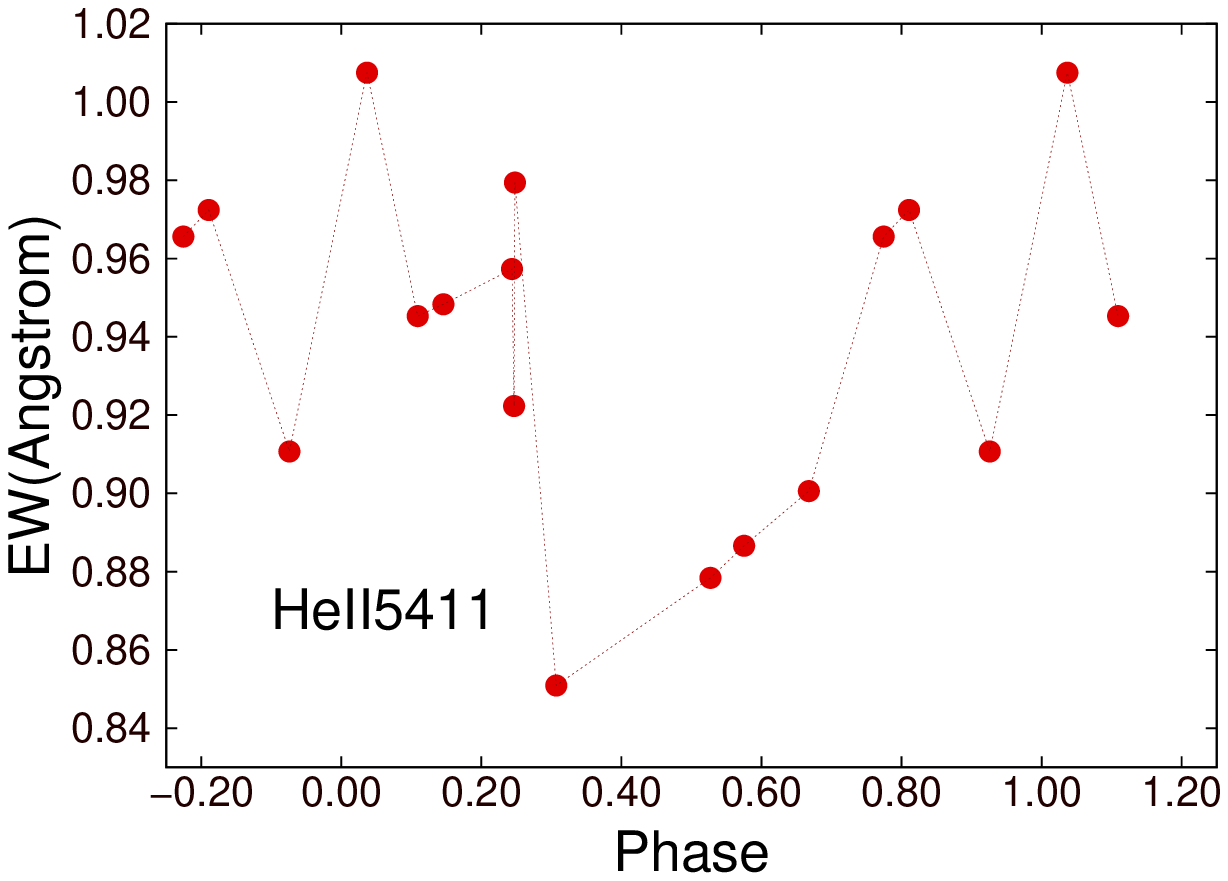}
\includegraphics[width=0.22\textwidth]{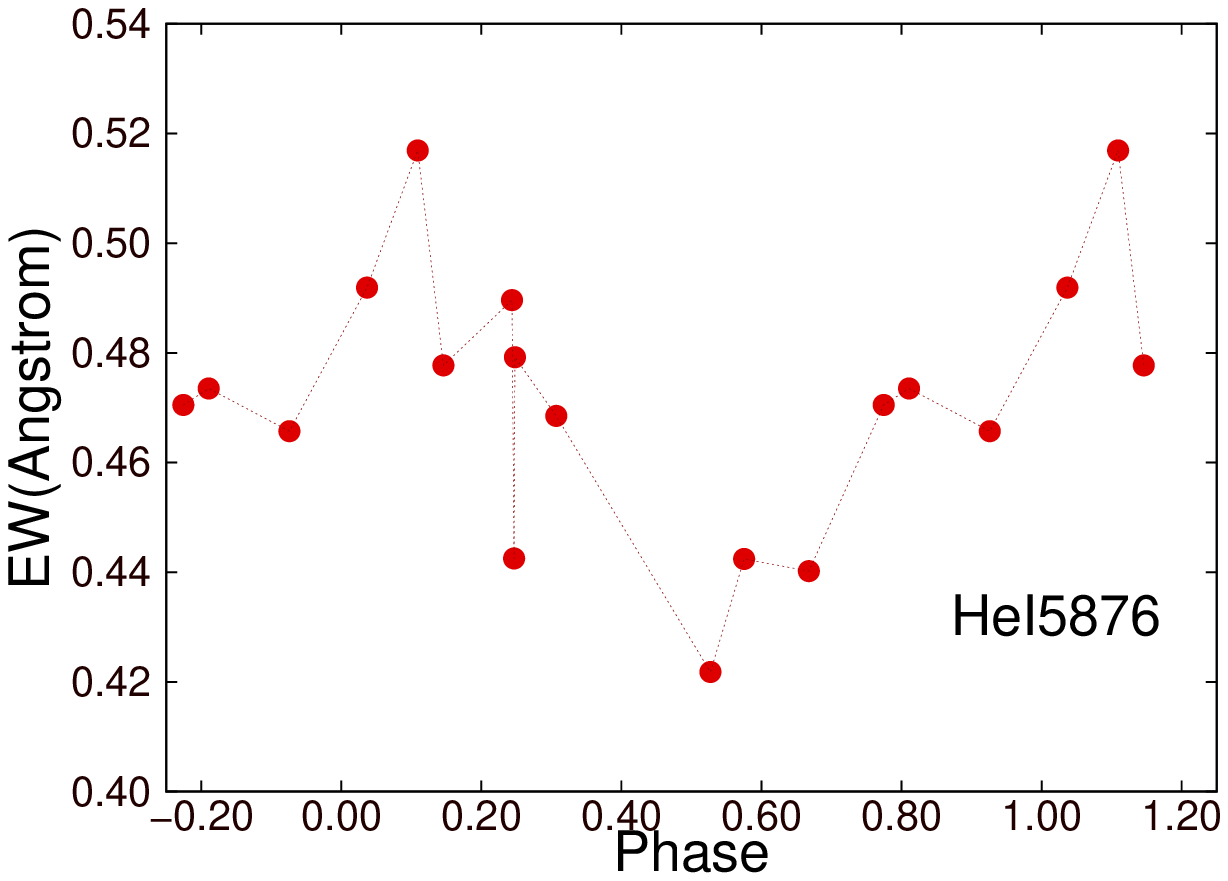}
\includegraphics[width=0.22\textwidth]{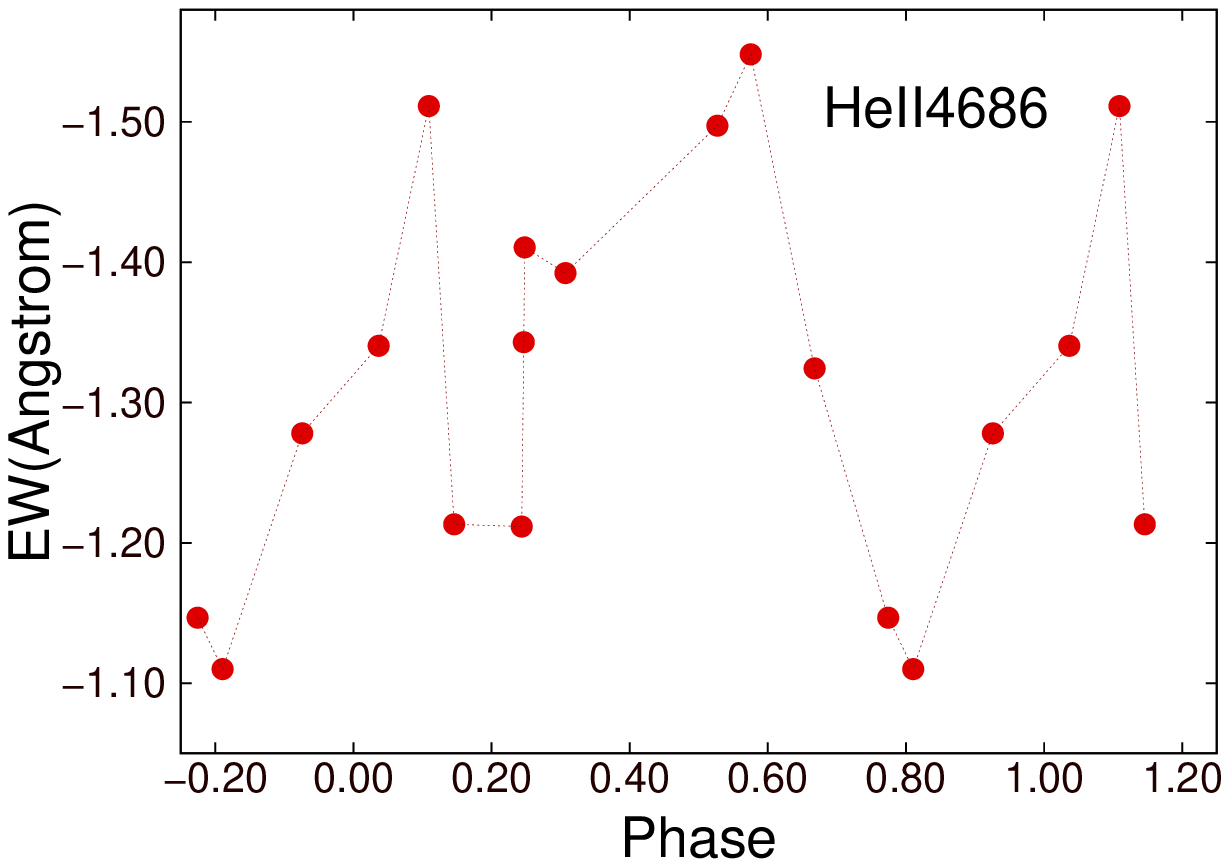}
\includegraphics[width=0.22\textwidth]{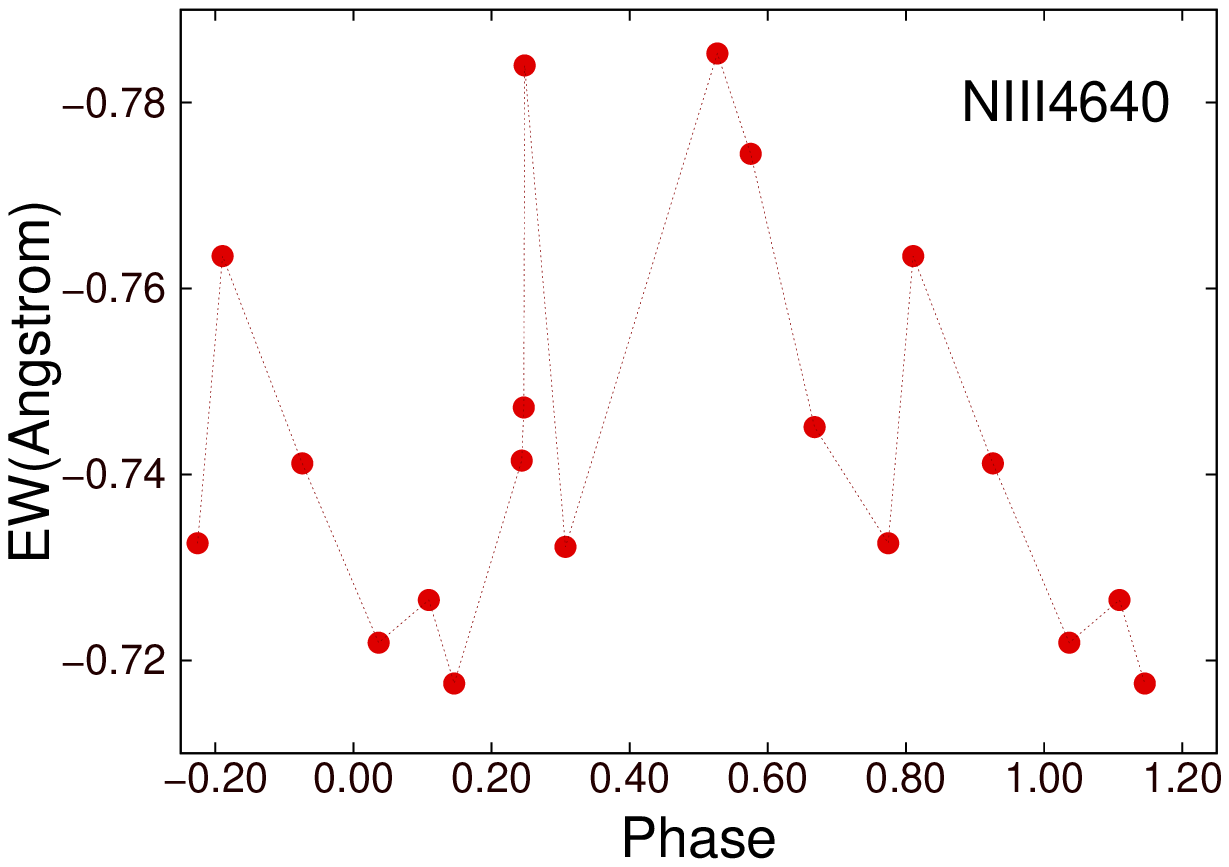}
\caption{
Variability of equivalent widths of lines belonging to different elements over the period of 1.78\,d.
The measurement accuracy corresponds to the size of the plotted symbols.
}
\label{fig:ew}
\end{figure}

\begin{figure}
\centering
\includegraphics[width=0.22\textwidth]{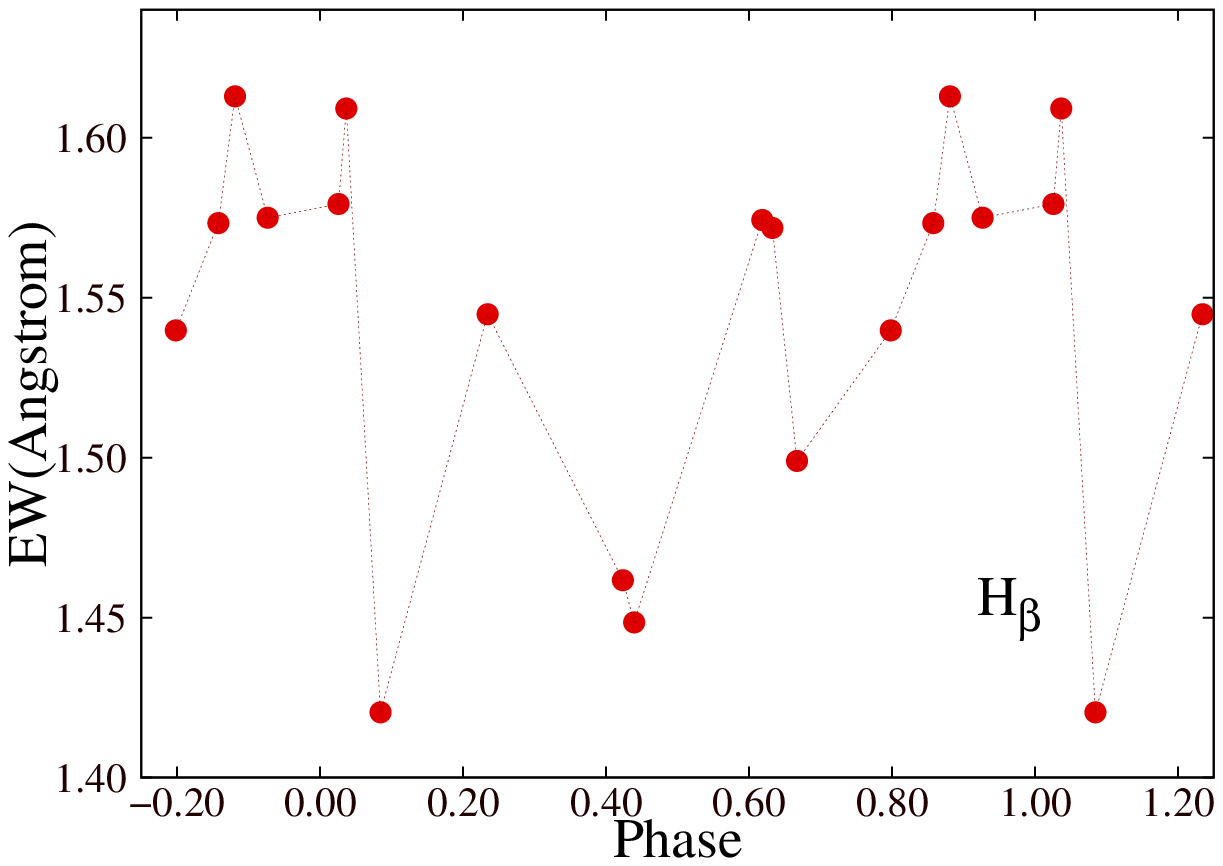}
\includegraphics[width=0.22\textwidth]{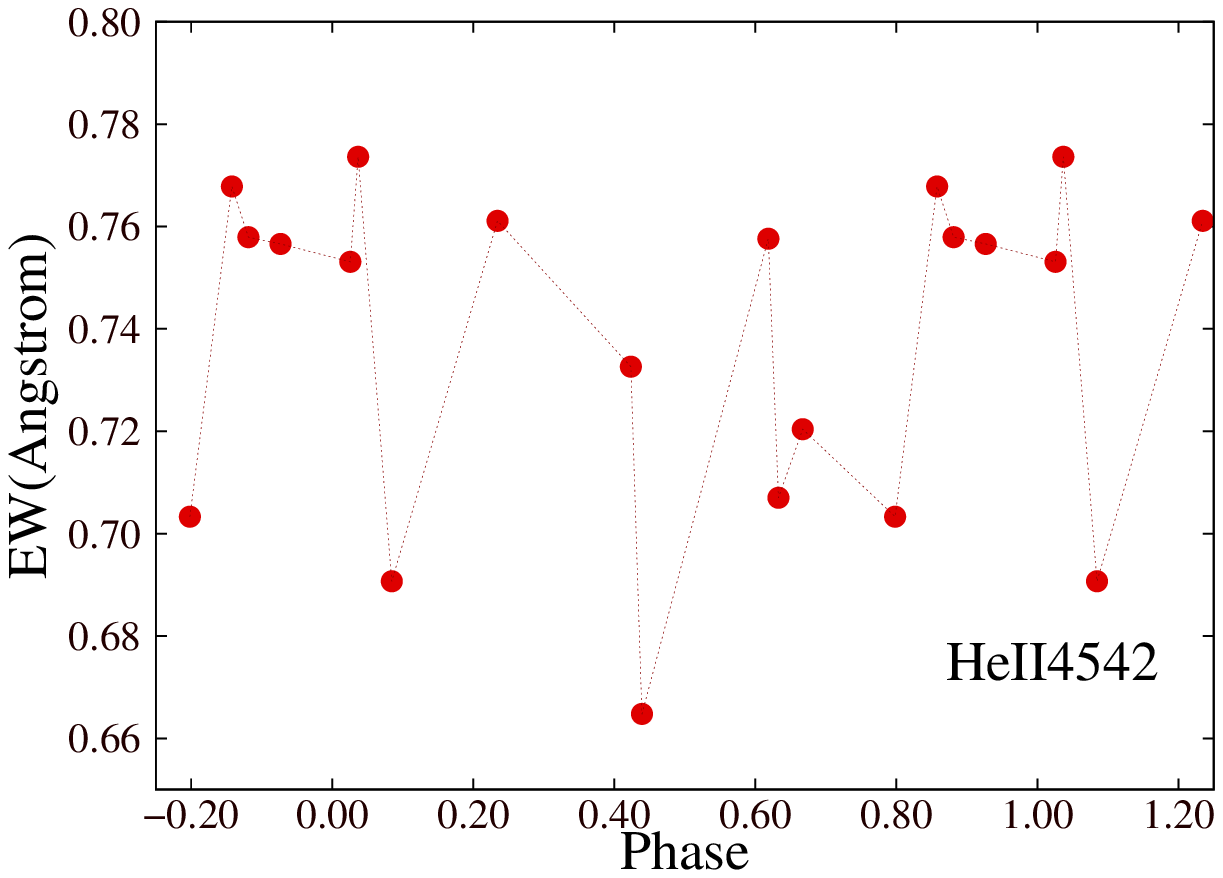}
\includegraphics[width=0.22\textwidth]{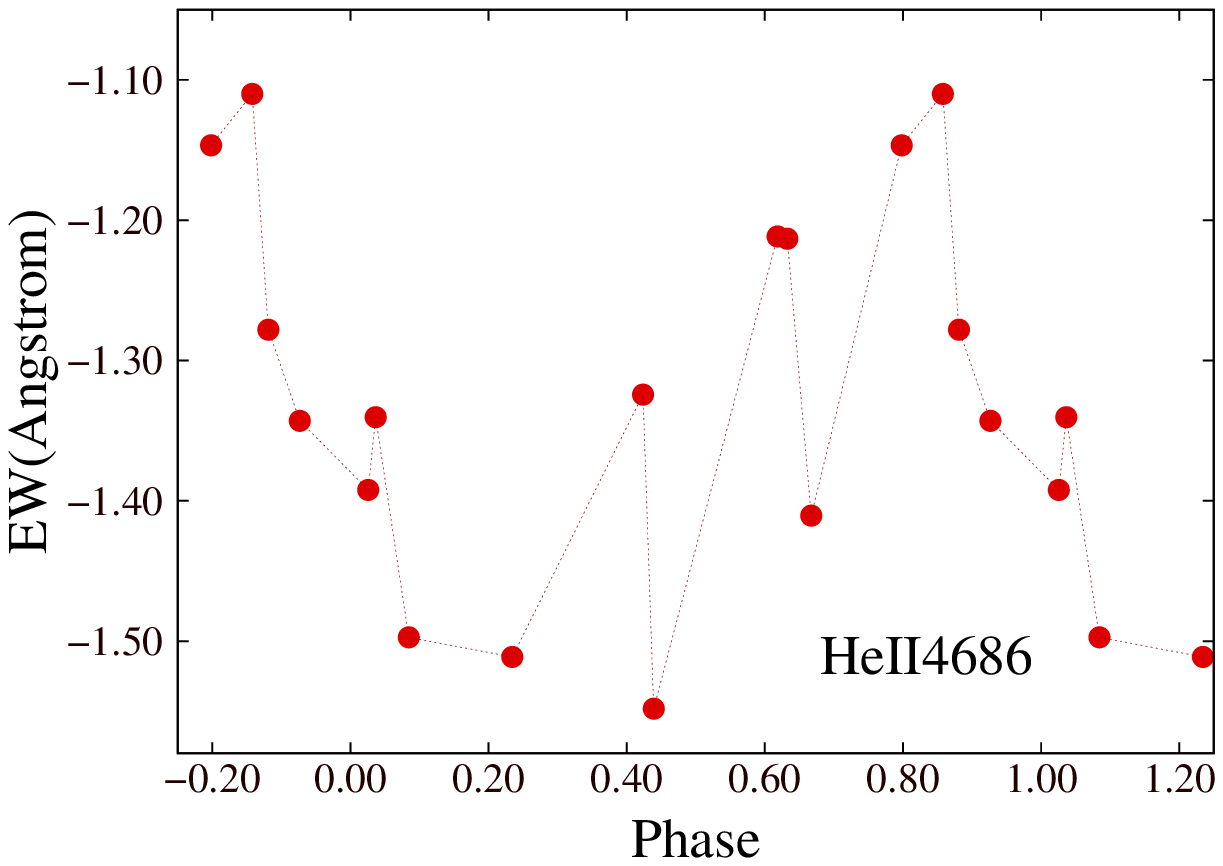}
\caption{
Variation of equivalent widths phased with a period of 5.26\,d of the lines 
H$\beta$, \ion{He}{ii}~4542, and \ion{He}{ii}~4686.
}
\label{fig:per5}
\end{figure}

\begin{figure}
\centering
\includegraphics[width=0.45\textwidth]{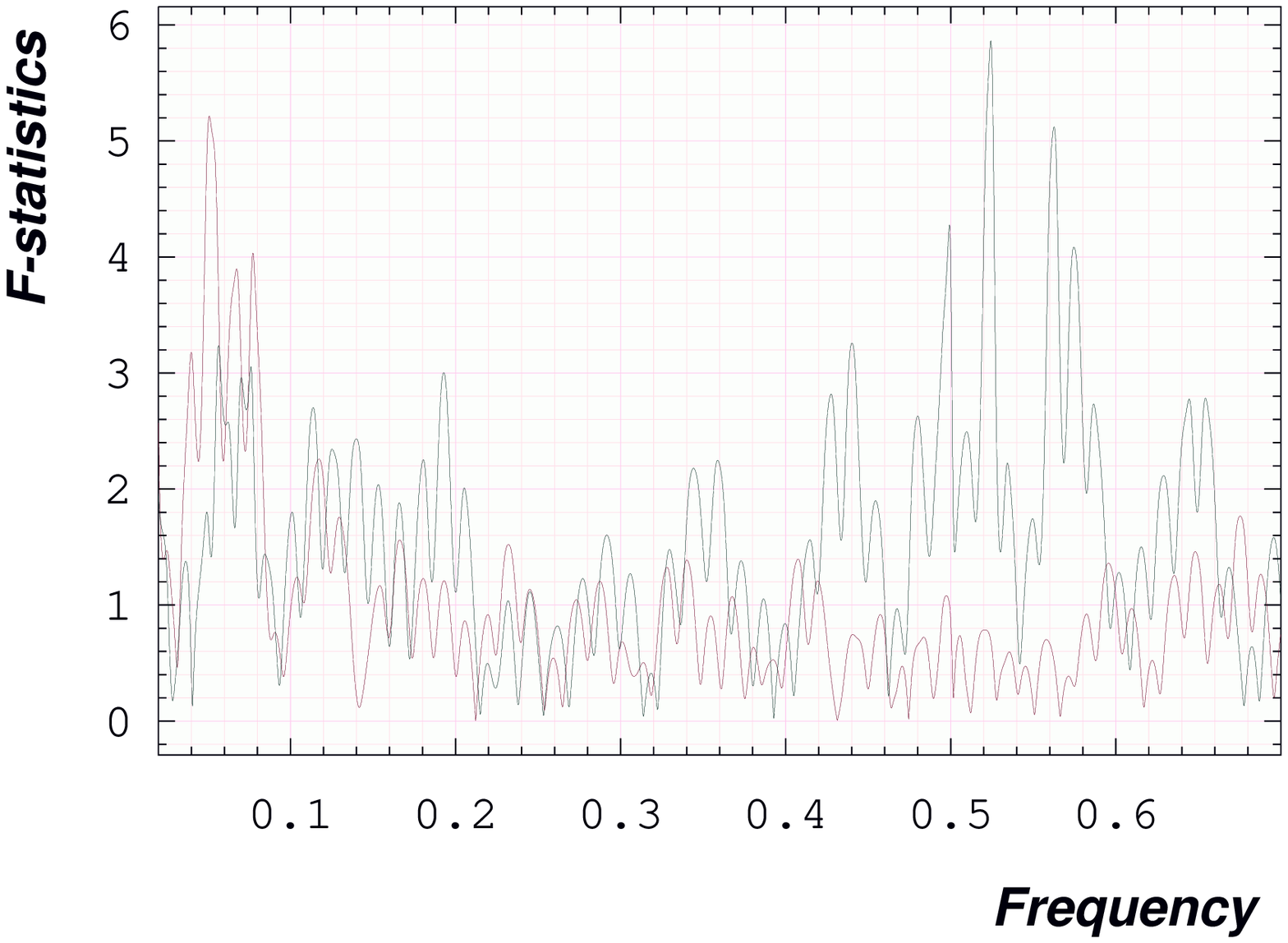}
\caption{
Frequency periodogram (in d$^{-1}$) for the equivalent width measurements for the H$\beta$ line.
The window function is indicated by the red color. 
}
\label{fig:per_eq}
\end{figure}

It is intriguing that the presence of a period of 1.78\,d is indicated in the variability of equivalent widths of 
lines belonging to different elements.
In Fig.~\ref{fig:ew}, we present their variations over the period of 
1.78\,d. Clear variability is detected for all 
studied absorption lines. The equivalent widths of absorption hydrogen and helium lines decreases in 
the vicinity of the positive magnetic pole, while the intensity of the
emission lines \ion{He}{ii}~4686 and the blend of \ion{N}{iii} lines at $\sim$4640\,\AA{} behave opposite, 
showing an increase at the positive magnetic pole. However, as is presented in Fig.~\ref{fig:per5}, no smooth variation 
of equivalent widths over a longer period  (we use in these plots the period of 5.26\,d suggested by \citealt{bal1992})
is detected.
Our frequency periodogram obtained for equivalent width measurements of H$\beta$ is presented 
in Fig.~\ref{fig:per_eq}.  The second largest peak in this periodogram is close to 0.56\,d$^{-1}$ (equivalent to 
a period of 1.78\,d) while the largest peak corresponds to a period of about 1.91\,d.
Interestingly, a smaller peak is found close to 0.192\,d$^{-1}$,
corresponding to a period of about 5.21\,d.

Further, our study of the temporal behaviour of different lines shows a shift in the line position at phases 
between 0.80 and 0.90 by about 60\,km\,s$^{-1}$ where the change from positive to negative polarity of the 
longitudinal magnetic field takes place. No such strong shift is detected at the phase when the field changes the polarity from 
negative to positive.
The \ion{He}{ii}~4686 line shows the presence of two components, where the blue component 
becomes stronger in the vicinity of the positive magnetic extremum and slightly stronger again in the vicinity of the negative field
extremum.
The \ion{He}{ii}~4542 line shows some kind of double-peak structure, where
the intensity of each absorption peak changes with time. However, more spectra are needed to understand the character of the 
variability in more detail. Noteworthy, similar  double-peak structures are observed in the shapes of He lines 
in the spectra of the Of?p star  CPD\,$-28^{\circ}$\,2561 \citep{Hubrig2015a} and the Wolf-Rayet star WR\,6 
(Hubrig et al., submitted to MNRAS). 

\section{Discussion}
\label{sect:disc}

No magnetic field is detected in $\zeta$\,Pup, as no magnetic field measurement has a 
significance level higher than 2.4$\sigma$. 
However, we cannot exclude a possible single-wave variation of the longitudinal magnetic field measurements.
The period of 1.78\,d detected by \citet{How2014} appears convincing and significant, and was 
also confirmed by BRITE observations (Ramiaramanantsoa, priv.\ comm.).
However, \citet{How2014} proposed that this period may be explained by stellar pulsations, while
our data suggest that the period may be explained by stellar rotation.
It is important to confirm this suggestion,
using more extensive spectroscopic material.

Our work indicates that spectroscopic/spec\-tro\-po\-la\-ri\-me\-tric studies appear most suitable for the determination of
rotation periods.
For instance, X-rays from massive stars, although present, are not well understood.
The search of periodical modulation in X-ray observations of $\zeta$\,Pup has a long history, but no conclusive 
detection was presented in any study yet.  
\citet{berg1996} used {\em ROSAT} to monitor $\zeta$ Puppis
over eleven days, totaling 56\,ks of observing time. Simultaneously with the X-ray observations,
the variability in the H$\alpha$ line was also monitored. The authors reported a 16.667\,h modulation both in the 
H$\alpha$ line as well as in the X-ray band pass between 0.9 and 2.0\,keV. 
However, this periodicity was not confirmed by \citet{osk2001} and  \citet{naze2013}, who 
used extensive {\em ASCA} and {\em XMM-Newton} observations.
Instead, some modulations with an amplitude of $\sim 15$\%\ on a time scale longer than 1\,d were found,
but these did not show coherent, systematic periodicity.

Summarizing, our search for the presence of a magnetic field in the fast rotating runaway stars $\zeta$~Oph and $\zeta$\,Pup
did not result in any significant detection.
Admittedly, a major part of the magnetic massive 
stars rotate rather slow, but still a small group of early B-type stars with strong magnetic fields and extraordinary fast 
rotation exists, posing a mystery for theories of star formation and magnetic field evolution 
(e.g.\  \citealt{riv2010}; \citealt{Hubrig2015b}).
The origin of magnetic fields in massive stars is still under debate. It was suggested that
the fields are either ``fossil'' remnants of the Galactic ISM field, which is amplified
during the collapse of a magnetised gas cloud (e.g.\ \citealt{price2007}), or that 
they are formed in a dramatic close binary interaction,
i.e., in a merger of two stars or a dynamical mass transfer event (e.g.\ \citealt{fer2009}).
\citet{Tetzlaff2010} reinvestigated the scenario of a binary SN in Upper Scorpius
involving $\zeta$\,Oph and PSR~B1929+10 and concluded that it is very likely that both objects were ejected during the 
same supernova event. Although in the case of $\zeta$\,Oph binary interaction is expected, no significant field
was detected in this star.
On the other hand, the results of our previous studies seem to indicate that the presence
of a magnetic field  is more frequently detected in candidate runaway
stars than in stars belonging to  clusters or associations (e.g.\ \citealt{Hubrig2011c}).
 Since the number of detected magnetic O-type stars is still rather small, these results need to be confirmed 
with a larger sample in the future.

\acknowledgments
Based on observations obtained in the framework of the ESO Prg.\ 092.D-0209(A).
AK thanks SPBGU for grant 6.38.18.2014.

\appendix




\end{document}